\documentclass{iopart}

\usepackage{epsf}
\usepackage[xdvi]{graphicx}

\begin{document} 

\flushbottom

\setcounter{page}{1}



\letter{Absorbing phase transition in a conserved lattice gas
with random neighbor particle hopping}

\author{S. L\"ubeck and A. Hucht}

\address{
Theoretische Tieftemperaturphysik, 
Gerhard-Mercator-Universit\"at,
Lotharstr.\,1, 47048 Duisburg, Germany\\[2mm]
Received 23 March 2001}

{\vspace{-7.0cm}
\noindent
accepted for publication in {\it J.\,Phys.\,}A
\vspace{6.5cm}}

\begin{abstract}
A conserved lattice gas with random neighbor
hopping of active particles is introduced which
exhibits a continuous phase transition from an active 
state to an absorbing non-active state. 
Since the randomness of the particle hopping breaks 
long range spatial correlations our model mimics the
mean-field scaling behavior of the recently introduced
new universality class of absorbing phase transitions
with a conserved field.
The critical exponent of the order parameter 
is derived within a simple approximation.
The results are compared with those of simulations
and field theoretical approaches.
\end{abstract}




\nosections

Recently Rossi\,\etal introduced a conserved
lattice gas (CLG) with a stochastic short range 
interaction that exhibits a continuous phase transition
to an absorbing state at a critical value of the
particle density~\cite{ROSSI_1}.
The CLG model is expected to belong to the universality
class of absorbing phase transitions with a 
conserved field~\cite{ROSSI_1}. 
This universality class is different from the
well known universality class of directed percolation
(see~\cite{JANSSEN_1,GRASSBERGER_2}, or for an 
overview~\cite{HINRICHSEN_1}).
In this work we introduce a modified CLG model with
a random neighbor interaction.
This random neighbor interaction suppresses 
long range correlations and the model
is expected to be characterized by the mean 
field scaling behavior.
The critical exponent of the
absorbing phase transition is derived within 
a simple approximation.
Numerical simulations of various lattice types
with different numbers of next neighbors 
confirm the obtained results.

Let us consider a system consisting of $L$ sites on
a chain with periodic boundary conditions.
At the beginning one distributes randomly $N=\rho L$ 
particles on the system where $\rho$ denotes the
particle density.
A particle is called active if at least one of its two
neighboring sites is occupied.
In the original CLG model active particles jump in
the next update step to one of their empty nearest 
neighbor site, selected at random~\cite{ROSSI_1}.
In the steady state the system is characterized by the
density of active sites $\rho_{\rm a}$ which
depends on $\rho$.
The density of inactive sites is given by $\rho-\rho_{\rm a}$ 
and $1-\rho$ is the density of empty sites.
The density $\rho_{\rm a}$ is the order parameter of the absorbing
phase transition~\cite{ROSSI_1}, i.e., it vanishes at the 
so-called critical density of particles $\rho_{\rm c}$
(which of course is lower than the trivial value $\rho=1/2$).

In our modification of the CLG model active particles 
are moved to a randomly chosen empty lattice site
which suppresses long range correlations.
On the other hand short range correlations still do exist.
For instance close to the critical point, where
the density of active sites is sufficiently low,
the density-density correlation function displays
an alternating behavior due to the repulsive interaction
of the particles.
But these correlations are short ranged 
and it is therefore possible to neglect them.
The values of the critical exponents are not affected by this
approximation.

Let us consider a configuration $\cal{C}$ of the 
lattice with $n$ active particles.
The number of active sites may change in each 
particle hopping, i.e., in each elementar update step.
For instance if both new neighbors of the hopped
particle are empty the number of active particles
is reduced by one, $\Delta n =-1$.
Without correlations the corresponding probability 
is $(1-\rho)^2$.
If one of the new neighbors of the hopped particle 
is occupied by an inactive particle ($p=\rho-\rho_{\rm a}$)
and the second neighbor is empty ($p=1-\rho$), 
the number of active sites is increased by one ($\Delta n =1$), 
and the corresponding probability is given by 
$p=2  (1-\rho) (\rho-\rho_{\rm a})$.
All other possible configurations and the corresponding
probabilities are listed in table\,\ref{table:conf}.

\begin{table}[b]
\caption{\label{table:conf}The configuration of the lattice before ($\cal{C}$) and
after ($\cal{C'}$) a particle hopping.
Only the target lattice site where a particle hops and its
left and right neighboring sites are shown.
Empty sites are marked by $\circ$, inactive sites are
marked by  $\ast$, and active sites by $\bullet$.
$\Delta n$ denotes the change of the number of active
sites due to the particle hopping and $p$ is the
corresponding probability of the configuration $\cal{C}$
if one neglects spatial correlations.}
\begin{indented}
\item[]\begin{tabular}{ccrl}
\br
\quad$\cal{C}$ & \quad$\cal{C'}$       & \quad$\Delta n$       & \quad$p(\cal{C})$ \\  
\mr \\
\quad$\circ \circ \circ $\quad     &  \quad$\circ \ast \circ $\quad         & \quad$-1$\quad 	& \quad$(1-\rho)^2$\quad   \\ 
\quad$\ast  \circ \circ$      	   &  \quad$\bullet \bullet \circ$     	    & \quad$+1$      	& \quad$2 (1-\rho)(\rho-\rho_{\rm a})$   \\ 
\quad$\ast  \circ \ast $      	   &  \quad$\bullet \bullet \bullet$   	    & \quad$+2$ 	& \quad$(\rho-\rho_{\rm a})^2$   \\ 
\quad$\bullet \circ \circ$         &  \quad$\bullet \bullet \circ$     	    & \quad$0$  	& \quad$2 \rho_{\rm a} (1-\rho)$   \\ 
\quad$\bullet \circ \bullet$       &  \quad$\bullet \bullet \bullet$   	    & \quad$0$  	& \quad$\rho_{\rm a}^2$   \\ 
\quad$\bullet \circ \ast $         &  \quad$\bullet \bullet \bullet$   	    & \quad$+1$ 	& \quad$2 \rho_{\rm a}(\rho-\rho_{\rm a})$   \\ 
\br
\end{tabular}
\end{indented}
\end{table}

\begin{figure}[t]
\begin{center}
\includegraphics[width=7cm,angle=0]{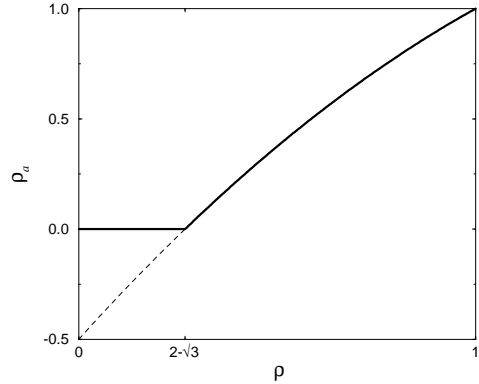}
\end{center}
\caption{\label{fig:rho_a_mf}Figure 
    The density of active sites $\rho_{\rm a}$ as a function of the 
    particle density~$\rho$ in the steady state.
    Negative values (dash line) correspond to an absorbing state
    with $\rho_{\rm a}=0$.}
\end{figure}

In this way one can calculate the probabilities
that the number of active particles are changed 
by $\Delta n$ and one gets 
\begin{equation}
\begin{array}{lcl}
p_{\scriptscriptstyle \Delta n = -1} & =  &(1-\rho)^2\\ \\
p_{\scriptscriptstyle \Delta n = 0}  & =  & 2 \, \rho_{\rm a} \, (1-\rho) \, +\, \rho_{\rm a}^2\\ \\
p_{\scriptscriptstyle \Delta n = 1 } & =  & 2 \, (\rho-\rho_{\rm a}) \, (1-\rho) \, 
+\, 2\, \rho_{\rm a} \, (\rho-\rho_{\rm a})\\ \\
p_{\scriptscriptstyle \Delta n = 2 } & =  & (\rho-\rho_{\rm a})^2\\
\end{array}
\end{equation}
The expectation value of $\Delta n$ is 
\begin{equation}
E[\Delta n] \; = \; \sum_{\Delta n=-1}^{2} \, \Delta n \, \,
p_{\scriptscriptstyle \Delta n} 
\; = \; -1 -2\rho_{\rm a} + 4 \rho -\rho^2.
\label{eq:expect_value}
\end{equation}
The average number of active sites is constant in the
stationary state, i.e., the expectation value
of $\Delta n$ should be zero in the steady state.
Using the constraint $E[\Delta n]=0$ it is possible
to calculate $\rho_{\rm a}$ as a function of $\rho$
and one gets
\begin{equation}
\rho_{\rm a} \;  = \; \frac{\,4 \rho- \rho^2 -1\,}{2}.
\label{eq:rho_a}
\end{equation}
The corresponding curve is plotted 
in figure\,\ref{fig:rho_a_mf} for $0<\rho<1$.
Negative values of $\rho_{\rm a}$ corresponds
to an absorbing state (i.e.~$\rho_{\rm a}=0$).
The critical point is determined by $\rho_a=0$, which
leads to $\rho_{\rm c} = 2 -\sqrt{3}$ (the second solution 
can be neglected since $\rho_{\rm c}=2+\sqrt{3}>1$).
Writing $\rho_{\rm a}$ as a function of the 
reduced density $\delta\rho = \rho- \rho_{\rm c}$
one gets
\begin{equation}
\rho_{\rm a} \; = \; \sqrt{3} \, \delta\rho \; 
-\frac{1}{2} \, \delta\rho^2.
\label{eq:rho_a_scal}
\end{equation}
Thus the order parameter of the absorbing phase
transition vanishes in leading order as
$\rho_{\rm a} \sim \delta\rho$, i.e., the 
order parameter exponent is $\beta=1$.

We briefly remark that it is straight forward to 
generalize the above derivation from a chain with
two neighbors to a $d$-dimensional cubic lattice with
$z=2d$ neighbors.
Since $\Delta n$ depends on the number of 
inactive sites ($\ast$) only (see table\,\ref{table:conf})
the corresponding probabilities are just polynomials
in $\rho_\ast=\rho-\rho_{\rm a}$, i.e.,
\begin{equation}
\begin{array}{lcl}
p_{\scriptscriptstyle \Delta n = -1} & =  &(1-\rho)^z,\\ \\
p_{\scriptscriptstyle \Delta n = 0}  & =  & (1-\rho_\ast)^z \, - \, (1-\rho)^z,\\ \\
p_{\scriptscriptstyle \Delta n{\scriptscriptstyle \ge} 1}      & =  & 
{z \choose {\Delta n}}\, \rho_\ast^{\Delta n} \, (1-\rho_{\ast})^{z-\Delta n}.\\
\end{array}
\end{equation}
Using again the steady state condition $E[\Delta n]=0$
one gets
\begin{equation}
\rho_{\rm a} \;  = \; \rho \, - \, \frac{(1-\rho)^z}{z}.
\label{eq:rho_a_z}
\end{equation}
The critical density $\rho_{\rm c}$ is determined by $\rho_{\rm a}=0$.
Expanding equation\,(\ref{eq:rho_a_z}) around $\rho_{\rm c}$ yields
\begin{equation}
\rho_{\rm a} \; = \; (2-\rho_{\rm c}) \; \delta\rho \,
+ \, \Or(\delta\rho^2),
\label{eq:rho_a_scal_z}
\end{equation}
i.e., the order parameter exponent is $\beta=1$ independent
of the number of next neighbors.

\begin{figure}[t]
\begin{center}
\includegraphics[width=7cm,angle=0]{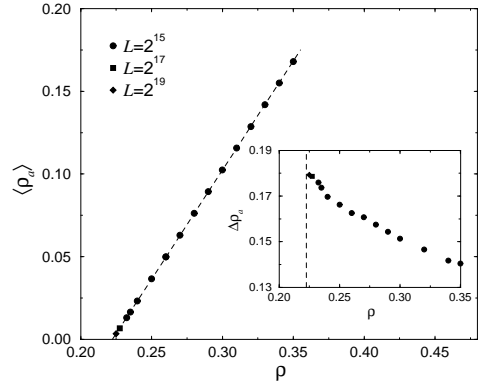}
\end{center}
\caption{\label{fig:rho_a_z2}Figure
    The average density of active sites for a one-dimensional chain
    of size~$L$ with random particle hopping ($z=2$).
    The dashed lines corresponds to a linear fit, i.e., the
    order parameter exponent is $\beta=1$.
    The inset displays the fluctuations of the
    order parameter which exhibits a discontinuous
    behavior at the critical density $\rho_{\rm c}$ (dash line).}
\end{figure}

\begin{figure}[b]
\begin{center}
\includegraphics[width=7cm,angle=0]{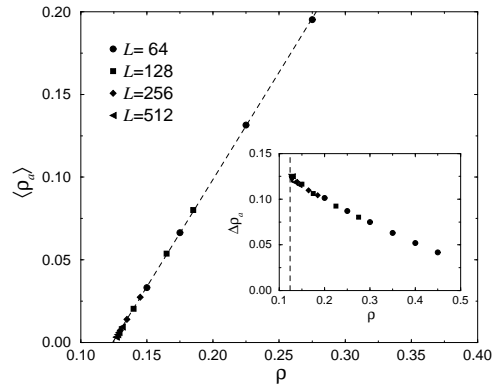}
\end{center}
\caption{\label{fig:rho_a_z4}Figure 
    The density of active sites for a two-dimensional square
    lattice of linear size~$L$ with random particle hopping ($z=4$).
    The dashed lines corresponds to a linear fit, i.e., the
    order parameter exponent is $\beta=1$.
    The inset displays the fluctuations of the order 
    parameter which exhibits a discontinuous
    behavior at the critical density $\rho_{\rm c}$ 
    (dash line).}
\end{figure}

In the following we compare this results with those
obtained from simulations.
We simulated a one dimensional chain ($z=2$) and a two
dimensional square lattice ($z=4$) where the active
particles are moved to a randomly chosen empty lattice
site.
In both cases a random sequential update was used
and the results are plotted in figure\,\ref{fig:rho_a_z2}
and figure\,\ref{fig:rho_a_z4}, respectively.
As expected the critical density $\rho_{\rm c}$ decreases 
with the number of next neighbors $z$.
Note that the critical value of the 
chain $\rho_{\rm c}\approx 0.2224$ and of the square
lattice $\rho_{\rm c}\approx 0.1244$ differs slightly 
from the above analytical result 
$\rho_{\rm c}=0.2679...$ and $\rho_{\rm c}=0.1380...$,
respectively.
This deviation, which decreases with increasing~$z$, 
is caused by the neglection of
correlations between neighboring sites.
Simulations reveal that these correlations exists but
are of short range (not shown).
Therefore, the critical value is shifted but the scaling 
behavior itself agrees with our analytical results, i.e.,
the order parameter vanishes at the critical point
continuously with an exponent $\beta=1$.

Additionally to the order parameter its fluctuations
\begin{equation}
\Delta \rho_{\rm a} \; = \; L^D \, \left (\langle \rho_{\rm a}^2 \rangle
\, - \, \langle \rho_{\rm a}\rangle^2 \right )
\label{eq:fluc_01}
\end{equation}
are measured in the simulations.
Here $D$ denotes the dimension of the system.
As can be seen from the insets of figure\,\ref{fig:rho_a_z2}
and figure\,\ref{fig:rho_a_z4} the fluctuations exhibit 
a discontinuous behavior (jump) at $\rho_{\rm c}$.

The value $\beta=1$ and the jump of the fluctuations
was also observed in the CLG model above the critical dimension 
$D_{\rm c}=4$~\cite{LUEB_19} where the scaling 
behavior of the model is determined by the mean-field
exponents.
Furthermore the values $\beta=1$ and $D_{\rm c}=4$ 
were predicted within a field theoretical
approach which is expected to represent the 
universality class of absorbing phase transitions
with a conserved field~\cite{PASTOR_2,PASTOR_4}.

Finally we just mention that the derivation of our results 
corresponds to a mapping of the dynamics 
to a branching process (see for instance~\cite{KANNAN_1}).
There each active site can create $i\in\{0,1,2,3\}$ 
active sites in the next generation with the 
probability $p_i=p_{\scriptscriptstyle \Delta n=i-1}$.
The steady state condition corresponds 
to the condition that the average number of created 
active sites in the next generation is one which 
yields directly equation\,(\ref{eq:rho_a}).

\nosections{S.\,L.~would like to thank H.\,K.~Janssen
for helpful discussions and useful comments on 
the manuscript.}




\end{document}